\documentclass[graybox, envcountchap]{svmult}
\usepackage{hyperref}
\usepackage{mathptmx}        % selects Times Roman as basic font
\usepackage{helvet}          % selects Helvetica as sans-serif font
\usepackage{courier}         % selects Courier as typewriter font
\usepackage{makeidx}         % allows index generation
\usepackage{graphicx}        % standard LaTeX graphics tool
\usepackage{multicol}        % used for the two-column index
\usepackage[bottom]{footmisc}% places footnotes at page bottom
\usepackage[utf8]{inputenc}
\usepackage{mathtools}
\usepackage{tcolorbox}
\usepackage{rotating}

\usepackage[nonumberlist,nopostdot,xindy,toc,acronym]{glossaries}
\setglossarystyle{altlist}

\newglossaryentry{coopetition}{
    name={coopetition},
    description={TODO}
}

\newglossaryentry{classification}{
    name={classification},
    description={A \gls{supervised-learning} task  in which labels are categorical variables (discrete, usually non ordinal).}
}

\newglossaryentry{regression}{
    name={regression},
    description={A \gls{supervised-learning} task in which labels are continuous or ordinal (categorical regression).}
}

\newglossaryentry{dataset}{%
  name={dataset},%
  description={A collection of labeled or unlabeled \gls{data-samples}.}%
  }

\newglossaryentry{dataset-requirements}{
    name={dataset requirements},
    description={The set of specifications of a dataset outlined in a document drafted before and collection begins. It should include, among other things, nature of the data, purpose, quality and quantity, means, costs, timeline.}
}

\newglossaryentry{dataset-design}{%
  name={dataset design},%
  description={The formal definition of a particular \gls{data-collection}.}.
  }
  
\newglossaryentry{dataset-implementation}{%
  name={dataset implementation},%
  description={Carrying out a particular \gls{data-collection} according to a given \gls{dataset-design}.}.
  }
  
\newglossaryentry{dataset-life-cycle}{
    name={dataset life-cycle},
    description={The set of processes which creates or transforms a dataset from \gls{dataset-requirements} to its \gls{dataset-distribution} and \gls{dataset-maintenance} or \gls{dataset-deprecation}.}
}

\newglossaryentry{data-collection}{
    name={data collection},
    description={The set of processes involved in producing a \gls{dataset}.}
}

\newglossaryentry{data-gathering}{
    name={data gathering},
    description={A particular \gls{data-collection}, which consists in gathering existing \gls{data-samples} or an entire existing \gls{dataset}, possibly from different sources, on which we do not have any influence.}
}

\newglossaryentry{data-reusing}{
    name={data reusing},
    description={The action of using an existing \gls{dataset} in an application, without changing its original purpose. Reusing is distinct from \gls{data-repurposing} or \gls{data-recycling}.}
}

\newglossaryentry{data-repurposing}{
    name={data repurposing},
    description={The action of using an existing \gls{dataset} in a new task or application, by changing its original purpose. Repurposing may include changing \gls{data-annotation}, re-sampling, transforming. Repurposing is distinct from \gls{data-reusing} or \gls{data-recycling}.}
}

\newglossaryentry{data-recycling}{
    name={data recycling },
    description={A particular \gls{data-collection}, which consists in recycling existing \gls{data-samples} or an entire existing \gls{dataset}, possibly from different sources, on which we do not have any influence. Recycling is distinct from \gls{data-reusing} or \gls{data-repurposing}.}
}

\newglossaryentry{data-resampling}{
    name={data resampling},
    description={The action selecting particular \gls{data-samples} in a \gls{dataset}. This may yield \gls{sample-bias} (or correct it).}
}

\newglossaryentry{synthetic-data-generation}{
    name={synthetic data generation},
    description={Is the process of generating artificial data that mimics real-world observations and can be used to (pre)train machine learning models when actual data is difficult or expensive to get.}
}

\newglossaryentry{data-bias}{
    name={data bias},
    description={The existence of spurious correlations (or dependencies) between data labels \gls{spurious-feature}s. Data bias includes \gls{sample-bias} and \gls{confounding-bias}}
}

\newglossaryentry{sample-bias}{
    name={sample bias},
    description={A particular type of \gls{data-bias} induced by selecting samples in a non-representative way of the target population.}
}

\newglossaryentry{confounding-bias}{
    name={confounding bias},
    description={A particular type of \gls{data-bias} induced by the existence of a common cause between labels and \gls{spurious-feature}s.}
}

\newglossaryentry{spurious-feature}{
    name={spurious feature},
    description={A variable (or feature or attribute) of \gls{data-samples}, which may be a ``protected attribute'' that could trigger discrimination (\eg gender, ethnicity, or age) or a variable known to be irrelevant to the particular task at hand (\eg the identity of the data collector, the time of data collection, the recording conditions).}
}

\newglossaryentry{core-feature}{
    name={core feature},
    description={A variable (or feature or attribute), which is though of as ``leggitimate'' to use to perform a \gls{learning-task}, as opposed to a \gls{spurious-feature}. Note that it is possible, because of \gls{entanglement} that some variable in a \gls{data-represention} combine original spurious and core features. Merely suppressing spurious features may not resolve the problem of bias in data.}
}

\newglossaryentry{data-simulation}{
    name={data simulation},
    description={A particular \gls{data-collection}, which consists in artificially creating data with a numerical simulator.}
}

\newglossaryentry{data-acquisition}{
    name={data acquisition},
    description={A particular \gls{data-collection}, which consists in measuring a real world signal with instruments, and converting it to digital information.}
}

\newglossaryentry{data-annotation}{
    name={data annotation},
    description={The action of attaching to existing data another type of complementary data, often called \gls{meta-data}, providing extra information (\eg object bounding boxes).}
}

\newglossaryentry{meta-data}{
    name={meta-data},
    description={Various \gls{data-annotation}s providing extra information to \gls{data-samples} (\eg object bounding boxes, date and time of collection, data origin).}
}

\newglossaryentry{data-labelling}{
    name={data labelling},
    description={A particular \gls{data-annotation}, which consists in mapping data samples to class labels. By extension, in the case of regression, labels can be continuous variables.}
}

\newglossaryentry{data-samples}{
    name={data samples},
    description={Individual units in a dataset.}
}

\newglossaryentry{data-transformation}{
    name={data transformation},
    description={A set of processes, which map an input data representation to an output data representation.}
}

\newglossaryentry{data-augmentation}{
    name={data augmentation},
    description={A set of processes which generates new data samples based on a fixed input dataset.}
}

\newglossaryentry{data-integration}{
    name={data integration},
    description={A set of processes which merge different datasets with complementary properties such as features, sources or quantities.}
}

\newglossaryentry{data-fusion}{
    name={data fusion},
    description={Synonym of data-integration.}
}

\newglossaryentry{data-cleaning}{
    name={data cleaning},
    description={A set of processes which detect, remove or replace mistakes in data samples.}
}

\newglossaryentry{data-reduction}{
    name={data reduction},
    description={A set of processes which reduce the quantity of information contained in a dataset for example to have improve computational efficiency or to reduce noise.}
}

\newglossaryentry{data-representation}{
    name={data representation},
    description={A well specified data structure in which each \gls{data-samples} is stored in a digital form in a computer. Feature vectors are a frequently used data representation in Machine Learning.}
}

\newglossaryentry{data-normalization}{
    name={data normalization},
    description={...}
}

\newglossaryentry{data-calibration}{
    name={data calibration},
    description={...}
}

\newglossaryentry{dataset-evaluation}{
    name={dataset evaluation},
    description={A set of processes, which verify that the produced \gls{dataset} meets the \gls{dataset-requirements}.}
}

\newglossaryentry{dataset-distribution}{
    name={dataset distribution},
    description={Means of making a \gls{dataset} available to the public.}
}

\newglossaryentry{dataset-maintenance}{
    name={dataset maintenance},
    description={Means of ensuring error corrections, upgrades, user-feedback, new regulations, or new ethical standards, are incorporated into new versions of the dataset.}
}

\newglossaryentry{dataset-retirement}{
    name={dataset retirement},
    description={Retiring a \gls{dataset} from public availability because of irreversible errors, new regulations, or new ethical standards.}
}

\newglossaryentry{dataset-deprecation}{
    name={dataset deprecation},
    description={Notification that a \gls{dataset} should not be used anymore but remains public for traceability or restricted usage because of irreversible errors, new regulations, or new ethical standards.}
}

\newglossaryentry{data-soundness}{
    name={data soundness},
    description={...}
}

\newglossaryentry{data-completeness}{
    name={data completeness},
    description={...}
}

\newglossaryentry{data-compactness}{
    name={data compactness},
    description={...}
}

\newglossaryentry{learning-task}{%
  name={learning task},%
  description={A formal prediction problem, denoted as $T$, composed of \gls{training-data} and \gls{validation-data} or \gls{test-data}, depending on the challenge phase, with the objective of (1) producing a \gls{predictor} using a \gls{trainer} (supplied by a challenge participant) using training data, and (2) evaluating the \gls{predictor}, using validation or test data. The evaluation is carried out using one (or several) \gls{evaluation-metric}s.}%
}

\newglossaryentry{evaluation-metric}{%
  name={evaluation metric},%
  description={A function $L(\hat{y}, y)$, of a prediction  $\hat{y}=f(x)$ made by a \gls{predictor} $f(.)$ on a input sample $x$ (from \gls{validation-data} or \gls{test-data}) and an associated ``ground truth'' value $y$. A score is obtained by averaging $L(\hat{y}, y)$ over all validation of test samples, which estimates empirically the generalization performance of the predictor. }
  }

\newglossaryentry{training-data}{%
  name={training data},%
  description={A subset of the \gls{dataset} $D_{tr} = \{ (x_i, y_i) \}_{i=1}^n \subseteq X \times Y$, which represent a snapshot of the ``real-world'', assumed to be drawn from a probability distribution $P$ on $X \times Y$, which represents the abstract process generating the data (usually unknown). This subset is used by the \gls{trainer} (or learning algorithm).}%
}

\newglossaryentry{validation-data}{%
  name={validation data},%
  description={A subset of the \gls{dataset}, disjoint from the \gls{training-data}, but usually drawn from the same distribution, used to get feed-back on performance during the development of a learning algorithm, so the \gls{test-data} remains untouched until the final evaluation. Using test data to adjust hyper-parameters of select algorithms results in optimistically biased performances. In a challenge, validation data are used for evaluation during the \gls{development-phase} (or feed-back phase).}%
}

\newglossaryentry{warmup-phase}{%
  name={warmup phase},%
  description={A challenge phase (also called public phase or challenge beta-testing) during which the participants are invited to try out the challenge protocol before the official starting date of the \gls{development-phase}, using some sample (public) data, not used in subsequent phases.}
  }

\newglossaryentry{development-phase}{%
  name={development phase},%
  description={A challenge phase (also called feed-back phase), during which the participants develop a solution to the problem of the challenge and submit results for immediate evaluation on \gls{validation-data}, to received feed-back on a (public) leaderboard.}
  }
  
  \newglossaryentry{final-phase}{%
  name={final phase},%
  description={A challenge phase (also called final test phase), during which the participants are evaluated on \gls{test-data}, not used either for training or validation. The final phase leaderboard remains private (only visible to the organizers) until the challenge is over.}
  }
  
\newglossaryentry{test-data}{%
  name={test data},%
  description={A subset of the \gls{dataset}, disjoint from the \gls{training-data} and the \gls{validation-data}, used to get feed-back on performance during the development of a learning algorithm, so the \gls{test-data} remains untouched until the final evaluation. Using test data to adjust hyper-parameters of select algorithms results in optimistically biased performances. In a challenge, test data are used for evaluation only once (for each final entry of each participant) in the \gls{final-phase}.}%
}

\newglossaryentry{alpha-challenge}{%
  name={$\alpha$-challenge},%
  description={}%
}

\newglossaryentry{beta-challenge}{%
  name={$\beta$-challenge},%
  description={}%
}

\newglossaryentry{predictor}{%
  name={predictor},%
  description={A function $f(.)$ mapping an input $x$ to an output $y$, solving a {classification} or a \gls{regression} problem. An \gls{alpha-challenge} asks participants to submit (trained) predictors.}%
}

\newglossaryentry{trainer}{%
  name={trainer},%
  description={An algorithm (usually called learning algorithm), which outputs a \gls{predictor}, or more generally a trained learning machine, which may be a data generator. The input to a trainer is \gls{training-data}. A \gls{beta-challenge} asks participants to submit trainers, and eventually tests on multiple \gls{learning-task}s their capability of automated machine learning (AutoML).}%
}

\newglossaryentry{meta-trainer}{%
  name={meta trainer},%
  description={An algorithm, which outputs a \gls{trainer}, given a dataset of datasets (meta-training set). A \gls{gamma-challenge} asks participants to submit meta-trainers, and tests on a meta-test set their capability of learning to learn (meta-learning).}%
}

\newglossaryentry{supervised-learning}{
    name={supervised learning},
    description={A \gls{learning-task} in which labels $Y$ are specified.}
}

\newglossaryentry{power-differential}{
    name={power differential},
    description={The difference in power between persons in positions of authority and those individuals in subordinate positions, which results in a vulnerability on the part of the subordinate. For example, the natural differences in power that exist between faculty and student.}
}

\newglossaryentry{observational-setting}{
    name={observational setting},
    description={A method for collecting data in which the investigator in charge of data collection does not interfere with the phenomenon. The distribution of samples collected should reflect the ``natural'' distribution of data.}
}

\newglossaryentry{experimental-setting}{
    name={experimental setting},
    description={A method for collecting data in which  the investigator  interferes with the natural world to achieve desired effects. A planned experiment consists in varying some factors systematically, in a controlled manner, or randomly.}
}

\newacronym{dlc}{DLC}{Data Life-Cycle}

\newacronym{ddlc}{DDLC}{Dataset Development Life-Cycle}

\newacronym{gdpr}{GDPR}{General Data-Protection Regulation}

\newacronym{mllc}{MLLC}{Machine Learning Life-Cycle}

\newacronym{sari}{SARI}{State Action Reward Information}

\newacronym{dream}{DREAM}{Dialogue for Reverse Engineering Assessment and Methods}

\newacronym{casp}{CASP}{Critical Assessment of protein Structure Prediction}

\newacronym{miccai}{MICCAI}{Medical Image Computing and Computer Assisted Intervention Society}
\makeglossaries

\usepackage{xspace}
\newcommand*{\eg}{e.g.\@\xspace}

\usepackage{pdfpages}
\usepackage{amssymb,graphicx,amsmath} %,subfigure}
\usepackage{times}
\usepackage{tikz}
\usepackage{soul}
\usepackage{color}
\usepackage{xcolor}

\usepackage{enumerate}
\usepackage{comment}

\definecolor{MyDarkGreen}{rgb}{0.17,0.46,0.25} 
\definecolor{MyDarkRed}{rgb}{0.88,0.22,0.21} 
\definecolor{MyDarkBlue}{rgb}{0.11,0.11,0.70} 
\definecolor{lightgray}{gray}{0.85}
\definecolor{blue}{rgb}{0.1216, 0.4667, 0.7059}
\definecolor{orange}{rgb}{1.0, 0.4980, 0.0549}
\sethlcolor{Dandelion}

\usetikzlibrary{arrows,shapes,plotmarks,positioning}
\usetikzlibrary{decorations.markings}

\tikzset{>=stealth'} 
\tikzstyle{graphnode} = 
   [circle,draw=black,minimum size=22pt,text centered,text
     width=22pt,inner sep=0pt] 
\tikzstyle{var}   =[graphnode,fill=white]
\tikzstyle{vardashed}   =[graphnode,draw=gray,fill=white]
\tikzstyle{obs}   =[graphnode,fill=black,text=white]
\tikzstyle{obsgrey}   =[graphnode,draw=white,fill=lightgray,text=black]
\tikzstyle{par}    =[graphnode,draw=white,fill=red,text=black] 
 \tikzstyle{crucial} =[graphnode,draw=white,fill=yellow,text=black] 
\tikzstyle{fac}   =[rectangle,draw=black,fill=black!25,minimum size=5pt]
\tikzstyle{facprior} =[rectangle,draw=black,fill=black,text=white,minimum size=5pt]
\tikzstyle{edge}  =[draw=white,double=black,very thick,-]
\tikzstyle{blueedge}  =[draw=white,double=blue,very thick,-]
\tikzstyle{rededge}  =[draw=white,double=red,very thick,-]
\tikzstyle{prior} =[rectangle, draw=black, fill=black, minimum size=
5pt, inner sep=0pt]
\tikzstyle{dirprior} = [circle, draw=black, fill=black, minimum
size=5pt, inner sep=0pt]
\usepackage{graphicx,subcaption} % support the \includegraphics command and options
\usepackage{tikz}
\usetikzlibrary{calc,arrows}
\usepackage{color, colortbl}
\usepackage{booktabs} % for much better looking tables

\usepackage{amsthm}
\usepackage{amssymb}% http://ctan.org/pkg/amssymb
\usepackage{array} % for better arrays (eg matrices) in maths
\usepackage{paralist} % very flexible & customisable lists (eg. enumerate/itemize, etc.)
\usepackage{verbatim} % adds environment for commenting out blocks of text & for better verbatim
\usepackage{makecell}
\usepackage{array,multirow}
\usepackage{mathrsfs}
\usepackage[T1]{fontenc}
\usepackage{varwidth}
\usepackage{hyperref}
\usepackage{epigraph}
\usepackage{here}
\usepackage{dsfont}
\usepackage{amsfonts}
\usepackage{pifont}

\usepackage{forest}

\usepackage[disable]{todonotes}

\newcommand{\isabelle}[1]{\todo[inline,color=orange!40]{#1 -- Isabelle}}

\newcommand{\walter}[1]{\todo[inline,color=green!20]{#1 -- Walter}}

\usepackage{eurosym}
\usepackage[graphicx]{realboxes}
\usepackage{pdflscape}

\usepackage[lined, ruled, boxed, linesnumbered, commentsnumbered]{algorithm2e}

\title{Challenge design roadmap}

\author{Hugo-Jair Escalante, Isabelle Guyon, Addison Howard, Walter Reade, S\'ebastien Treguer}

\institute{INRIA \email{streguer@gmail.com}}
\definecolor{LightCyan}{rgb}{0.88,1,1}
\definecolor{LightYellow}{rgb}{1,1,0.88}

\begin{document}
\setcounter{chapter}{1}

\maketitle

\label{chap:design}

{\large Preprint, to appear in book: \\ AI competitions and benchmarks: The science behind the contests. \\ Adrien Pavao, Isabelle Guyon, Evelyne Viegas, Eds.} \\ \\

\abstract*{Abstract}

\abstract{
Challenges can be seen as a type of game that motivates participants to solve serious tasks. As a result, competition organizers must develop effective game rules. However, these rules have multiple objectives beyond making the game enjoyable for participants. These objectives may include solving real-world problems, advancing scientific or technical areas, making scientific discoveries, and educating the public.
In many ways, creating a challenge is similar to launching a product. It requires the same level of excitement and rigorous testing, and the goal is to attract "customers" in the form of participants. The process begins with a solid plan, such as a competition proposal that will eventually be submitted to an international conference and subjected to peer review. Although peer review does not guarantee quality, it does force organizers to consider the impact of their challenge, identify potential oversights, and generally improve its quality.
This chapter provides guidelines for creating a strong plan for a challenge. The material draws on the preparation guidelines from organizations such as Kaggle\footnote{\url{http://www.kaggle.com/}}, ChaLearn\footnote{\url{http://www.chalearn.org/}} and Tailor\footnote{\url{https://tailor-network.eu/}}, as well as the NeurIPS proposal template, which some of the authors contributed to.
}

\section{Before you start} 

\label{Design}

\begin{figure}
    \centering
    \includegraphics[width=0.75\textwidth]{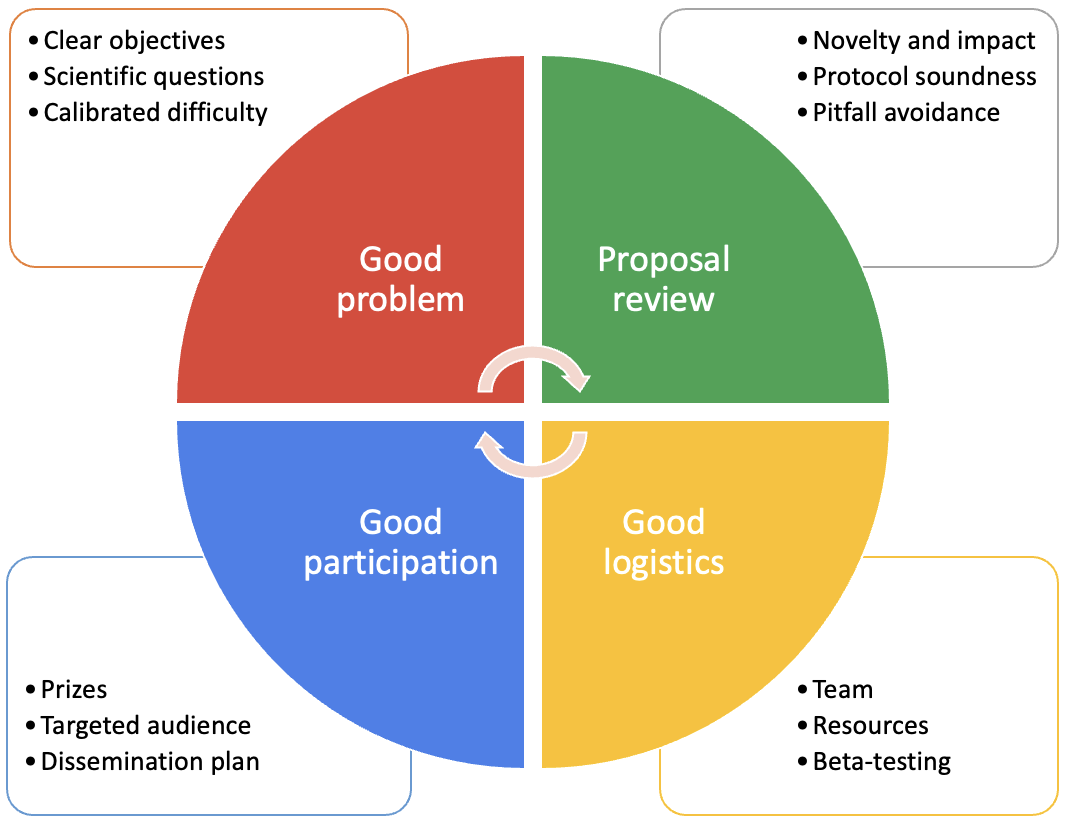}
    \caption{Challenge design principal pillars.}
    \label{fig:design}
\end{figure}

This section outlines the questions that challenge organizers need to ask themselves before starting the process of organizing a challenge. By doing so, organizers can avoid underestimating the resources they will need to achieve their goals and prepare adequately for their proposal. A schematic of the process is shown in Figure ~\ref{fig:design}.

\subsection*{Do you have a problem lending itself to a challenge?}

First of all, do you have a problem to be solved? Maybe not yet! You are just interested in becoming a challenge organizers. In that case, we recommend that you {\bf partner} with researchers, industrial or non-profit organizations who have data and an interesting problem to offer. In Table \ref{tab:datasets} we list some {\bf data sources} that can inspire you. Popular datasets can be good choices for benchmarking purposes. Using a popular dataset allows researchers to compare their results to a large body of prior work, which can provide context and establish a benchmark for future research. A dataset that has been used by many researchers in the past can be re-purposed to answer specific research questions. However, there are chances that the competitors (or the base models they are using, such as pre-trained deep-net backbones) may have been exposed to such data already, which could bias the evaluation and/or give an unfair advantage to some competitors. Hence it is recommended to draw on sources that have not yet been exploited by machine learning researchers. See the chapter on Data %\ref{chap:dataset} 
for useful tips on how to collect and prepare data.

\begin{table}[]
    \centering
    \begin{tabular}{|p{0.1\textwidth}|p{0.2\textwidth}|p{0.7\textwidth}|}
    \hline
    {\bf Domain} & {\bf Data source}     &  {\bf Data type}\\
    \hline
    \hline 
    \rowcolor{LightCyan}
    Machine Learning & \href{https://www.kaggle.com/datasets}{Kaggle datasets}     & The largest general-purpose ML dataset repository with >170K datasets in various formats, but generally coming with illustrative Jupyter notebooks. \\
    \hline 
    \rowcolor{LightCyan} CodaLab  & \href{https://codalab.org/}{CodaLab}     & A large repository with data from hundreds of competitions, mostly academic~\cite{pavao:hal-03629462}.\\
    \hline 
    \rowcolor{LightCyan}
     Machine Learning &\href{https://huggingface.co/docs/datasets/index}{Hugging Face}    & More than 1000 datasets with well defined format/metadata and data loaders. Moatly for Audio, CV, and NLP modalities.\\
    \hline 
    \rowcolor{LightCyan}
    Machine Learning & \href{https://www.openml.org/}{OpenML}     & More than 5000 datasets, mostly in tabular format. \\
    \hline 
    \rowcolor{LightCyan}
    Machine Learning & \href{https://archive.ics.uci.edu/ml/index.php}{UCI ML Repository} & Historcal repository created in 1987 by D. Aha and his students. Hundreds of datasets, mostly small and in tabular format. \\
    \hline   
    \rowcolor{LightCyan}
    Reinforcement Learning & \href{https://gymnasium.farama.org/}{Farama gymnasium} & New version of OpenAI gym. It includes a variety of environments, such as classic control problems and Atari games. \\
    \hline
    \hline
    \rowcolor{LightYellow}
     Miscellaneous & \href{http://data.gov}{Data.gov} & Over 200,000 datasets from the US government. The datasets cover a wide range of topics, from climate to crime. \\
    \hline   
    \rowcolor{LightYellow}
    Sensor Data  & \href{https://www.noaa.gov/}{NOAA} \href{https://www.opensignal.com/}{OpenSignal} \href{https://arrayofthings.github.io/}{Array of Things)} & Sensor data from a variety of sources, such as IoT devices, wearables, and environmental sensors, can provide rich information about the physical world. This data can be used to develop models for a wide range of applications, such as health monitoring, environmental sensing, and predictive maintenance. \\
    \hline    
    \rowcolor{LightYellow}
     Audio data& \href{http://www.cs.toronto.edu/~complingweb/data/spRUCE/spRUCE.html }{SpRUce} \href{http://millionsongdataset.com/ }{Million Song} \href{ https://www.robots.ox.ac.uk/~vgg/data/voxceleb/ }{VoxCeleb}  & Audio data, such as speech and music, is a rich source of information that can be used for a variety of applications, such as speech recognition, speaker identification, and music recommendation.\\
    \hline    
    \rowcolor{LightYellow}
     Textual data& \href{https://commoncrawl.org/ }{Common Crawl} \href{https://archive.ics.uci.edu/ml/datasets/reuters-21578+text+categorization+collection }{ Reuters News} \href{ https://nijianmo.github.io/amazon/index.html}{Amazon Reviews} & Textual data, such as news articles, social media posts, and customer reviews, is a rich source of information that can be used for a variety of applications, such as sentiment analysis, topic modeling, and natural language understanding. \\
    \hline    
    \rowcolor{LightYellow}
    Satellite imagery & \href{https://www.planet.com/ }{Planet Labs } \href{ https://www.esa.int/}{ European Space Agency} \href{ https://www.nasa.gov/}{ NASA} & Satellite imagery can provide high-resolution images of the Earth's surface, which can be used for applications such as land use classification, urban planning, and disaster response. \\
    \hline    
    \rowcolor{LightYellow}
     Financial data & \href{https://www.quandl.com/ }{ Quandl} \href{https://finance.yahoo.com/ }{Yahoo Finance } \href{https://fred.stlouisfed.org/ }{ FRED} & Financial data, such as stock prices, market trends, and economic indicators, can be used to develop models for predicting stock prices, identifying trading opportunities, and understanding economic trends.\\
    \hline    
    \hline

    \end{tabular}
    \caption{{\bf Sources of data.} There is no good challenge without good data. An important aspect is to find ``fresh data'' to reduce the risk that the participants have been exposed to the challenge data previously, and can have an unfair advantage. Datasets commonly used in ML research (such as those colored in cyan) can be used as illustrative examples in the starting kit. Public datasets (such as those colored in yellow) can be used for the development phase (public leaderboard), but it is preferable to use novel fresh data for the final phase (private leaderboard).}
    \label{tab:datasets}
\end{table}

Second, do you have a {\bf good definition of your problem} and have you tried yourself to solve it with some {\bf simple baseline method}? Do you have a sense of how hard it is (it should {\bf neither be trivial nor too hard to solve})? If not, it is premature to organize and challenge with that problem, you need first to get familiar with this problem and be able to define criteria of success (called ``metrics''), and have some preliminary ideas on how to optimize them. Make sure you understand how to {\bf cast your problem into AI tasks}, which may range from machine learning tasks (binary classification, multi-class or multi-labels classification, regression, recommendation, policy optimization, etc.) \cite{burkov2019hundred},  optimization tasks (continuous, combinatorial or mixed; single- or multi-objective, etc.) \cite{alma9926520161805776}, reasoning (logic or probabilistic), planning, constraint programming, \ldots \cite{russel2010}, or a combination of several type of tasks. Having participated yourself to a challenge may be helpful. In that respect, the Kaggle book provides a gentle introduction to data competitions~\cite{banachewicz2022kaggle}.

Third, not all problems lend themselves to a challenge. Importantly, challenges are ``games of skill'', NOT ``games of chance''. Can you devise a quantitative way of evaluating participants (using your metrics) in which {\bf chance plays no role} (this is a legal requirement to organize scientific contest in most countries, to avoid that they fall under gambling regulations)? This may be particularly difficult. If the evaluation of participant entries relies on a statistic computed from data (typically some test set performance score using your metric), do you have {\bf enough data} to obtain small error bars or a stable ranking of participants? In \cite{guyon1998size}, the authors provide a rule-of-thumb for classification problems.
 
Usually, tens or thousands of examples must be reserved for test data, if you want good error bars; alternatively, you can use multiple datasets. Remember that, if the participants are allowed to make multiple submissions to the challenge and get feed-back on their performance on the test set on a leaderboard, they will soon overfit the leaderboard. In statistics, this is known as the problem of multiple testing. Leaderboard overfitting can be alleviated, to some extent, with a technique called ``The ladder" \cite{pmlr-v37-blum15}, which essentially boils down to quantizing the test set performance. However, to really prevent leaderboard overfitting, it is advisable to run the challenge in {\bf multiple phases}: during a development phase, the participants are allowed to make multiple submissions and view their performances on a ``public leaderboard''; during the final phase, the winners are chosen on the basis of a single submission, evaluated on a separate fresh test set. The performances are kept secret on a so-called ``private leaderboard'', invisible to the participants until the challenge is over. Since challenges are run in this way, leaderboard overfitting seems to have largely disappeared \cite{NEURIPS2019_ee39e503}. 

Be mindful that the metric really reflects the objective you want to optimize: a common mistake, for instance, is to use the error rate for classification problems, not distinguishing between the cost for false positive and false negative errors (e.g., it may be more detrimental to send a sick patient back home than to ask a sound patient to do one more medical exam). Also, would you rather your model provide you with a definitive prediction or a range of confidences? 
Finally, will you be declaring ties if performances between two participants are too close? %See Chapter \ref{chap:judging} to get ready to answer these last two questions. 

\isabelle{Walter TODO: Rewrite this paragraph}
On another note, data quantity is not the only concern, do you have {\bf quality data}? Can you guarantee, for instance, that data are free from biases? If you are not familiar with {\bf``data leakage}\footnote{\url{https://www.kaggle.com/docs/competitions\#leakage}}''\cite{Perlich}, consult Appendix C. If you know nothing about bias, learn about it \cite{https://doi.org/10.1002/widm.1356}, and beware of recycling or re-purposing data if you do not have detailed information about the original intent \cite{NEURIPS_DnB_2021_3b8a6142}. Can you guarantee that the {\bf test data is ``fresh''}, i.e., that none of the participants  had prior access to these data? Or, at least, if you are using public re-purposed data, that the data are {\bf obfuscated enough} that they are not recognisable to the participants, or hidden to the participants at test time?

And do not forget the check whether you have the {\bf right to use the data and/or code} that you need to use in your challenge!

\subsection*{What are your scientific questions?}

\label{ScientificQuestion}
Wait a minute, what is the main problem we want to address and would like to be solved?
Asking the good questions is key to get results inline with the initial goals.
What are the objectives of the challenge? Is our priority to address scientific questions, and which ones precisely, or to get as outcomes models easy to transfer to a production system with all its constraints in terms of robustness, explainability, performance monitoring, maintenance?
Is the only objective the final accuracy at the end of training without constraints on resources: compute, memory and/or time?
Or should the applicants also take into account limits in training time, computer power, memory size, and more, with the goal to find the sweet-spots for good trade-offs?

The definition of each task to achieve must help to solve a specific question raised by the challenge, but must also carefully take into account all constraints and reflections previously mentioned. 

Then what are the constraints in terms of data: volume, balance or unbalance of classes, fairness, privacy, external vs internal, etc.?
These questions are related to the tasks that are themselves related to the initial questions to be addressed. For each scientific question, you will then need to define some metrics allowing to measure how well each participant answers the question. %: More details in Chapter \ref{chap:judging}.

In general, AI competitions should have very specific objectives. There is a natural human tendency, when expending significant time and resources collecting and preparing data for a competition, to want to answer as many questions as possible from the competition. This is almost always counterproductive. While there may be considerable secondary information that can be gleaned at the conclusion of a competition, competitions should be designed to have a very specific primary question to be addressed. This primary question is commonly in the form of the maximum predictive performance a model can extract from a given dataset.

This primary question should be expressed in a single, objective metric. It is tempting, but generally unwise, to try to combine different metrics into a single objective. When disparate scientific questions are force-fit into a single evaluation metric, the unintended consequences may be that the setup favors optimization of one question over the others, or that all the questions sub-optimized to optimize the combined metric. However, there are scenarios where multiple objectives genuinely arise, necessitating consideration of several metrics. For example, the metrics might include both accuracy and speed or memory footprint. In such cases, it is essential to strike a balance, ensuring that all objectives are addressed without overly compromising on any single aspect. Properly defined and carefully weighted metrics can help ensure that all objectives are optimized without undue sacrifices.

When facing multiple primary questions, it can be beneficial to introduce separate competition tracks. For instance, the M5 forecasting competition featured a track for precise point forecasts\footnote{\url{https://www.kaggle.com/competitions/m5-forecasting-accuracy}} and another for estimating the uncertainty distribution of the values\footnote{\url{https://www.kaggle.com/competitions/m5-forecasting-uncertainty}}. Despite using the same dataset, the former adopted the weighted root mean squared scaled error, while the latter employed the weighted scaled pinball loss as their respective metrics. By creating distinct tracks, the competition enabled researchers to advance the state-of-the-art in each area, rather than settling for methods that performed adequately across both but didn't excel in either. The choice boils down to whether the aim is to nurture specialist or generalist algorithms. To motivate participants to engage in multiple tracks while still promoting comprehensive methods, an incentivized prize structure can be adopted. For example, winners might receive a reward of $x$ for one track, double that amount for two, and exponentially more, as in $2^n x$, for triumphing in $n$ tracks.

There is a competition format that does allow for more open-ended discovery of a dataset, which we'll refer to as an analytics competition. The idea here is to provide data, give general guidance on the objective, and allow the competitors to analyze the data for new insights. An example of this was the NFL Punt Analytics Competition\footnote{\url{https://www.kaggle.com/competitions/NFL-Punt-Analytics-Competition}}, where the goal was to analyze NFL game data and suggest rules to improve player safety during punt plays. While the scientific question was specific (``what rule changes would improve safety?"), the format allowed for a much broader exploration of the data and provided insights that wouldn't have surfaced by optimizing a single metric. While this can be beneficial, Analytics competitions tend to have much lower participation than predictive competitions, and require a significant amount of work after the competition deadline to review and manually score (using a predefined rubric) each of the submissions from the teams.

Useful guidelines have been provided by Frank Hutter in \href{https://slideslive.com/38923481/a-proposal-for-a-new-competition-design-emphasizing-scientific-insights}{``A Proposal for a New Competition Design Emphasizing Scientific Insights"}: ``The typical setup in machine learning competitions is to provide one or more datasets and a performance metric, leaving it entirely up to participants which approach to use, how to engineer better features, whether and how to pretrain models on related data, how to tune hyperparameters, how to combine multiple models in an ensemble, etc. The fact that work on each of these components often leads to substantial improvements has several consequences: (1) amongst several skilled teams, the one with the most manpower and engineering drive often wins; (2) it is often unclear {\bf why} one entry performs better than another one; and (3) scientific insights remain limited.
Based on my experience in both participating in several challenges and also organizing some, I will propose a new competition design that instead emphasizes scientific insight by dividing the various ways in which teams could improve performance into (largely orthogonal) modular components, each of which defines its own competition. E.g., one could run a competition focusing only on effective hyperparameter tuning of a given pipeline (across private datasets). With the same code base and datasets, one could likewise run a competition focusing only on finding better neural architectures, or only better preprocessing methods, or only a better training pipeline, or only better pre-training methods, etc. One could also run multiple of these competitions in parallel, hot-swapping better components found in one competition into the other competitions. I will argue that the result would likely be substantially more valuable in terms of scientific insights than traditional competitions and may even lead to better final performance.''

Machine learning challenges often aim to address a multitude of scientific questions. These questions can be categorized using a taxonomy based on their overarching 5W themes: what, why, how, whether, and what for. Here's a potential breakdown:
\begin{enumerate}

\item {\bf What (Discovery):} What patterns can be discovered in  data? What features are more significant or relevant to the target variable?
What groups or segments naturally form in the dataset? What are the characteristics of these clusters? An example of discovery challenge would be the \href{https://www.kaggle.com/c/higgs-boson/data}{Higgs Boson challenge}, that aimed at discovering a new high energy particle \cite{adam2015higgs}.

\item {\bf Why (Causality):}
Why did a specific event or outcome occur?
Are there variables that directly influence this outcome?
Why does a certain data point deviate from the norm?
For example, ChaLearn organized several \href{https://www.causality.inf.ethz.ch/challenge.php}{causality challenges}, including the Causation and Prediction challenge \cite{guyon2008design} and the Cause-effect pair challenge \cite{guyon2019cause}.

\item {\bf How (Prescriptive):}
How can we allocate resources efficiently to achieve a goal?
How can an agent take actions in an environment to maximize some notion of cumulative reward?
For example the \href{https://bbochallenge.com/}{Black box optimization challenge} asked participants to figure out how to optimize hyperparameters of models in a black box setting. The Learning to Run a Power Network (\href{https://l2rpn.chalearn.org/}{L2RPN}) challenge series is asking how an agent can control the power grid to transport electricity efficiently, while maintaining equipment safe \cite{marot2020learning,marot2020l2rpn,marot2021learning}.

\item {\bf Whether (Comparative):}
Whether Algorithm A is better than Algorithm B for a specific task? Whether a given preprocessing or hyperparameter setting X improves the model's performance over technique Y?
Whether there is a trade-off e.g., between performance metrics (like precision and recall).
Whether a model trained on dataset A performs better on dataset B compared to a model directly trained on dataset B (a transfer learning problem)?
Whether a certain RL method performs better in environment condition X compared to condition Y?"
For example, the Agnostic Learning vs. Prior Knowlege (\href{https://www.agnostic.inf.ethz.ch/}{AlvsPK}) challenge answers the question whether prior knowledge is useful to devise a good preprocessing or whether using agnostic features is enough \cite{guyon2007agnostic}.

\item{\bf What For (Purpose-driven):}
For whom might this model be unfair or biased?
For what populations does this model perform sub-optimally?
For what new tasks or domains can knowledge from one domain be useful?
For example, the \href{https://www.kaggle.com/c/jigsaw-unintended-bias-in-toxicity-classification}{Jigsaw Unintended Bias in Toxicity Classification} asked predicting and understanding toxicity in comments while considering potential biases against certain populations.
\end{enumerate}

This taxonomy not only provides a structured way to think about scientific questions in machine learning but also helps in deciding the kind of algorithms, data processing techniques, and validation strategies that might be best suited to address them.

\subsection*{Are you sufficiently qualified to organize your challenge?}

There are many difficult aspects to tackle in the organization of a challenge and this can be daunting. It is rare that a single person is qualified to address them all. Think of {\bf partnering with other experts} if you do not know how to answer any of the questions of the previous sections (or the following ones).

For instance, depending on the data types or modalities (tabular, univariate or multivariate time-series, image, video, text, speech, graph) and application-dependent considerations, 
{\bf appropriate evaluation metrics} should be chosen to assess performance of the submissions. It may make a lot of difference if one chooses accuracy rather than balanced accuracy or AUC if classes are imbalanced, for instance. You may know that and know the difference between MSE and MAE for regression, but do you know what SSIM, SHIFT, SURF are for image similarity? Do you know what F1 score is and what the difference is between micro-averaging and macro-averaging? Have you thought about whether you rather evaluate best objective value or time to reach given precision, for optimization tasks? Success or failure, time-to-solution for reasoning tasks? or do you need qualitative metrics possibly implying human evaluation (i.e., by some expert committee)? %For details, see Chapter \ref{chap:judging}.

Also, regarding data modalities, the choice of data and the evaluation of its quality require a lot of expertise. We have mentioned already the problem of data leakage, which leads to inadvertent disclosure of information about the ``solution'' of the challenge. The chapter on data %(Chapter \ref{chap:dataset}) 
reviews many more aspects of data that require attention, including legal aspects of ownership and attribution, privacy, fairness, and other legal aspects. Each aspect may be better handled by an appropriate expert.

Another aspect requiring expertise will be the {\bf preparation of baseline methods}. Make sure to include in your organizing team members who are knowledgeable of state-of-the-art methods and capable of implementing or running them on your tasks, to obtain baseline results. The code for baseline methods could be provided to the participants as part of a ``starting kit''. One motivating factor for the participant is ``upskilling'' themselves. Make sure you document well the baseline methods and provide good tutorial material, adapted to your audience. This will be much appreciated!

The adage "a rising tide lifts all boats" aptly fits this context. It conveys that when there's a general advancement or progress in a particular scenario (here referring to publicly accessible notebooks), all individuals involved reap the benefits, irrespective of their initial conditions. For instance, on Kaggle, you can kickstart a project using a pre-existing notebook created by someone else, paving the way for collective growth and assistance for all participants.

%More about building your team can be found in the introductory Chapter. % \ref{chap:introduction}.

\subsection*{Do you know your target audience?}
%\isabelle{I move this here ``as is''; did not review it yet.}

 It is important to define the target audience, in order to design a challenge which is attractive enough and adapt the level of difficulty with a barrier to enter that is not too high.
 However if the target audience is a mix of beginners and more experienced practitioners in Artificial Intelligence, a crucial issue is to find a sweet spot, to set the barrier low enough to allow for beginners to enter without too much headache, while keeping the competition challenging enough for experienced practitioners.
 Lowering the barrier to enter can be achieved by providing good documentation along with a simplified tutorial in a starting kit, providing  compute resources to make it accessible to anyone, not only people with own access to farms of GPU or TPU. And at the same time, keeping the problem to solve interesting enough for experienced practitioners might require several levels of difficulty, several phases of the challenge.

Choose the start date, time length, and time investment required, according to the targeted audience. If you target researchers, make sure that being successful doesn't involve too much engineering time efforts with respect to the scientific contribution, and coordinate with other conferences, workshops and other competitions in the field.
  
 Choose carefully a subject that is interesting for your targeted audience at the time of the competition.
 Problems may be unsuitable to make a good challenge not only for technical reasons, but also because it will either not raise sufficient interest or because it will raise eyebrows. Remember that prizes are only a small part of the incentive, because most participants do not win and the prizes usually are a very small compensation compared to the time spent: they are largely {\bf donating their time}! 

Ask yourself:
\begin{itemize}
\item whether you can illustrate your problem in one or several {\bf domains of interest to the public}, which may include medicine, environmental science, economy, sociology, transport, arts, education;
    \item whether the outcome of the challenge will have a {\bf practical societal or economical impact};
    \item whether the approached task (or any aspect associated to it, e.g., application, domain, scenario) represents a potential hazard to users' rights, including ethical issues, privacy concerns, etc.;
    \item how you can lower the barrier of entry to increase participation (winning when there are no other contestants is not fun!); this may mean e.g., using a uniform data format, providing data readers, providing sample code;
\end{itemize}

When introducing the topic of your challenge, it is essential to craft a compelling ``hook" that piques interest. Aim to attract a diverse range of participants. The beauty of this approach is that a machine learning expert, even without domain-specific knowledge, might clinch victory in a very specialized challenge. Conversely, an industry engineer looking to enhance their skills could very well triumph in a machine learning challenge.

Publication venue is an essential motivation for participants of academic challenges. However, a recent analysis has determined that prizes are the greatest factor in boosting competition for Kaggle participation.  While overall participation is mostly driven by the approachability (a competition with a tabular dataset will have more participation than one with 3D images), all else equal, the prize is 75 percent more important than any other factor. However, substantial prizes can attract participants more interested in monetary gain than scientific advancement, potentially leading to rule violations or the exploitation of challenge loopholes without disclosure.

\subsection*{What are your objectives?}

Are you more interested in finding a champion, benchmarking algorithms to evaluate or (incrementally) push the state of the art, or discovering brand new methods? Or are you simply interested in making your company/institution visible? While these three goals may be not mutually exclusive, your challenge design should take them into account. 

In {\bf recruiting challenges}, your goal is to find a {\bf champion}. You may want to select a representative problem of what it is like to work at your company or institution to find top talents to employ. You will NOT need to put in effort in preprocessing data, designing a good API, etc.: the participants will be expected to do all the dirty work and show they {\bf excel at solving all aspects of the problem}. Make part of the challenge deliverable that top ranking participants must deliver a technical report on their work to best evaluate them. 

In {\bf Research and Development challenges}, your goal is to benchmark algorithms. You may want to carefully design your API and have participants supply an object or a function, which addresses the {\bf specific sub-task you have identified as a bottleneck of your problem}. Make sure to sort out licensing conditions of the winner's solutions. One simple way is to ask winners to open-source their code as a condition of being eligible for prizes.

In {\bf Academic challenges}, your goal is to discover new methods for a problem that you largely do not know how to solve. You may want to have both quantitative and qualitative evaluations, e.g., in the form of a {\bf best paper award}. This is because it is not obvious when you invent a new method to optimize it and get it to outperform others quantitatively, this may involve tuning and engineering. 

In {\bf Public Relation challenges}, your goal is to make your company or institution known, e.g., to attract new customers or students, and to expose your specific data or problem of interest to this public. It is essential to keep the challenge as simple as possible (at least in its problem statement) and as didactic as possible, and build around it great communication with the public, including using mass media. You may want to have intermediate goals and prizes and organize interviews of participants making good progress, to boost participation.

In {\bf Branding challenges}, your goal is to put your name in front of a large community of data science practitioners by releasing a new technology or dataset (e.g., incentivizing the use of Tensorflow, or releasing a SOTA image classification set like ImageNet).

\subsection*{Do you have enough resources to organize a good challenge?}

Challenges with {\bf code submission} thought of being preferable to those with {\bf result  submission} in academia. This allows organizers to compare methods in a controlled environment and fosters fairer evaluations by providing to participants equal computational resources.
However, with the advent of large foundational models \cite{}, training on the challenge platform has become infeasible, for computational reasons. In that case, one can resort to letting the participants train their model on their own premises and submit the code of trained model, to be tested on the platform.
Also, while code competitions are often better to raise equity across participants without similar access to resources, computational constraints can hamper participants from using the processing pipelines they are used to, which would not lead to establishing state-of-the-art performance. Other elements of fairness may include not requiring entry fees or the purchase of material (e.g., robots), not to favor entrants who are economically advantaged.

Table \ref{tab:levels} illustrates a hierarchy of competition protocols for supervised learning tasks. As one progresses from one level to the next, participants have access to less information, whereas their submissions receive more. Level $\lambda$ is associated with challenges requiring result submissions, whereas the higher levels pertain to code submissions, from a class named ALGO. Level $\alpha$ is based on the premise of submitting pre-trained models. Level $\beta$  depicts a scenario of thorough code blind testing, where both training and testing occur on the platform. If multiple datasets are accessible, meta-testing can be conducted, given that participants are provided with examples of analogous tasks for meta-training. In the final stage, the $\gamma$ level, participants aren't provided with the meta-training set. Instead, they receive a basic task description.

Consequently, from level $\lambda$ to $\gamma$, the submissions evolve to be increasingly autonomous, aligning more closely with genuine automated machine learning (AutoML). However, this also necessitates increased computational resources to implement the challenge.

\begin{table}
    
\caption{\label{tab:levels} Hierarchy of competition protocols.}
{\footnotesize
\noindent
\begin{tabular}{|c|p{0.28\linewidth}|p{0.31\linewidth}|p{0.30\linewidth}|}

\hline
    {\bf Level} & {\bf Information available to participants }
     &  {\bf Information available to the algorithm only} &  {\bf Type of submission} \\
\hline
$\lambda$ & \cellcolor {blue!40} Everything, except test labels  &  Nothing & \verb|RESULTS| of test set predictions \\
\hline
$\alpha$ & \cellcolor {blue!20}  Labeled TRAINING set & \cellcolor {blue!10} Unlabeled TEST set  &  \verb|ALGO.predict()|  \\
\hline
$\beta$ & \cellcolor {blue!10} META-TRAINING set  &  \cellcolor {blue!20}  META-TEST set (each test set wo test labels) &  \verb|ALGO.fit()|, \verb|ALGO.predict()| \\
\hline
$\gamma$ & Nothing, except starting kit and sample data  & \cellcolor {blue!40} META-TRAINING set \& META-TEST set (each test set wo test labels)   &  \verb|ALGO.meta-fit()|, \verb|ALGO.fit()|, \verb|ALGO.predict()| \\
\hline

\end{tabular}
}
\end{table}

Do you have a {\bf budget} to cover these costs and others (like preparing data)? Here is a non-exhaustive list of possible costs. %See chapter \ref{chap:practical} for more details:

\begin{itemize}
    \item Data collection, labeling, cleaning.
    \item Data preprocessing and formatting.
    \item Compensation of engineers preparing baseline methods or impementing the challenge.
    \item Computational resources to run the challenge.
    \item Prizes.
    \item Advertising.
    \item Organization of a workshop.
    \item Travel awards to attend the workshop.
    \item Fees for a competition hosting service.
\end{itemize}

\subsection*{Do you have a plan to harvest the results of your challenge?}

Providing participants with opportunities to disseminate their work is both an important motivation to them to enter the challenge and a means of harvesting results. You may want to target one or several conferences (NeurIPS, KDD, WCCI, IDCAR have competition programs, others welcome competitions organized in conjunction with workshops). 

Let your participants know in advance when your competition will start, they will be looking forward to it. But, make sure you have enough time to get ready for the opening date! Avoid competitions having deadlines at the same time as conference deadlines or student exams. Recurring events can have a snowball effect: you progressively build a community, which progresses over the years to improve the state-of-the-art. 

Conferences do not usually have proceedings for their workshops, so you may have to make your own arrangements for proceedings. One venue that has been welcoming challenge proceedings in \href{https://proceedings.mlr.press/}{PMLR}, the Proceedings of Machine Learning Research. The recently founded \href{https://data.mlr.press/}{DMLR journal} has also been welcoming challenge papers.

At the very least, top ranking participants should be asked to fill out fact sheets (see and example in Appendix B). Fact sheets can be a mix of textual descriptions providing a brief report in human readable format, and survey answers to a few questions, which can easily be aggregated as statistics. It is best to ask the participants to fill out the fact sheets before revealing the final leaderboard results, because otherwise non-winners have little incentive to put in the effort. Also, it is best to put as a condition to winning prizes to open-source the code and fill out the fact sheets.

Do not under-estimate the duration of challenge preparation, which, depending on the data readiness, the complexity of implementation of the challenge protocol and of establishing  results may vary from a few days to over a year. Refer to the Chapter on datasets for recommendations on how to prepare data. 

\subsection*{Common pitfalls}
\label{sec:pitfalls}

Perhaps the most common mistake, is to offer a challenge that has a {\bf lack of clarity} in the problem definition and the goals to be reached, a {\bf too complex metric} defying intuition, or a {\bf lack of focus} by addressing too many problems at once. When designing an AI competition, it is important to understand that there is no way to optimize for all of the questions you might want to answer. It is better to put in the hard work up front to decide what the specific primary question should be and how to measure success with a single simple metric. If there are secondary questions that you would like addressed, these should only be considered if they can be answered without jeopardizing the primary question. Clarity in the {\bf rules} is also important. If you need a long list of rules for legal reasons, summarize them. Do not forget to have the participants accept the rules at the time of registration. See \href{http://www.causality.inf.ethz.ch/GeneralChalearnContestRuleTerms.html}{ChaLean contest rules}, for inspiration. Add a FAQ page answering most frequently asked questions.

\begin{figure}
    \centering
    \includegraphics[width=0.75\textwidth]{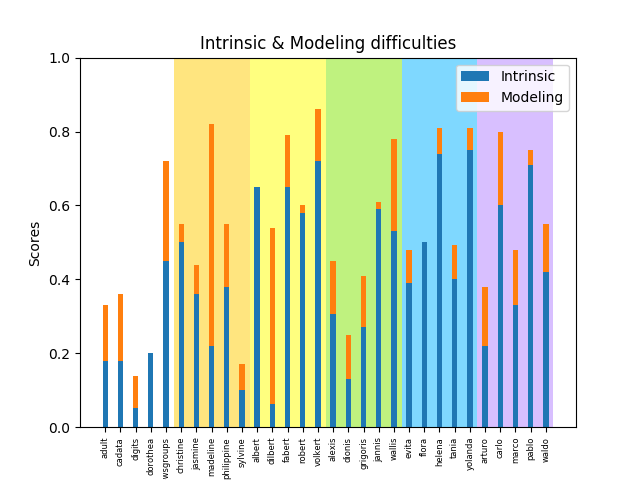}
    \caption{Example of intrinsic and modeling difficulty of datasets.}
    \label{fig:difficulty}
\end{figure}

Another pitfall is to discourage serious competitors to enter because of {\bf obvious flaws} (in data or in challenge rules). {\bf Beta-test} your challenge thoroughly using volunteers (who will then not be eligible to enter the challenge) or solicit people you know well to enter the challenge as soon as it opens and report possible flaws. If possible, make a {\bf ``dry run''} or organize first a scaled-down version of your competition to test its protocol, before you launch the ``big one''. Such trials will also allow you to {\bf calibrate the difficulty of the task} to the target audience. Figure \ref{fig:difficulty}, taken from the AutoML challenge \cite{guyon2019analysis}, shows how the difficulty of datasets might be calibrated. Test set scores are represented (normalized between 0 and 1, 0 is the level of random guessing). The height of the blue bar represents the score of the best model (our best estimated of the irreducible error of ``intrinsic difficulty''). The height of the orange bar represents the range of scores between the best and the worst model, which we use the evaluate the ``modeling difficulty''. The datasets that are most suitable to separate methods well are those with small ``intrinsic difficulty'' and large ``modeling difficulty''. During beta-testing, you may want to set up an ``inverted competition'' among organizers, in which datasets are submitted against established baseline methods, then select those datasets with the largest ratio of modeling difficulty over intrinsic difficulty. If after adjusting the task difficulty, there is still little participation during the challenge, be ready to {\bf lower the barrier of entry}, but providing more clues to get started, code snippets, notebooks, and/or tutorials.

Another way of discouraging participation is to put {\bf too many constraints} or prerequisites to enter the competition: having attended a previous challenge or event, registering with a nominative account, attending a conference, open-sourcing code, etc. While more constraints can be placed on the winners (or the entrants to the final phase), it is advisable to facilitate as much as possible entering the feed-back phase.

\isabelle{Walter TODO: Eliminate? This is maybe where leakage fits best.}
The success of the competition also rests on having {\bf quality data}, unknown yet to the public. We mentioned the problem of {\bf bias in data} or {\bf data leakage} in the introduction. Appendix C provides guidance on how to avoid falling into the most common traps when preparing data. In addition to having quality data, you must also have {\bf sufficiently large datasets}. A common mistake is to reserve a fixed fraction of the dataset for training and testing (typically $10\%$ of the data for testing), without anticipating the error bars. A simple rule-of-thumb to obtain at least one significant digit in a 1-sigma error bar, for classification problems, is to reserve at least $N=100/E$ test examples, where $E$ is the anticipated error rate of the best classifier (the challenge winner) \cite{guyon1998size}. So, if you anticipate $E=10\%$, $N=1000$, if you anticipate $E=1\%$, $N=10000$.

Last but not least: do not forget the {\bf incentives} (prizes, publication opportunities, etc.) and to {\bf communicate well with your participants}. This starts with announcing your challenge ahead of time and advertising aggressively a few days into the challenge (once you are confident everything is running smoothly). Use all possible means available: mailing lists, social media, personal contact. Monitor the level of participation, get feed-back and stimulate participation, if needed, by adding bootcamps, webinars, and tutorial sessions. Make use of a forum and stimulate discussions between organizers and participants and between participants.

\section{The proposal}

In the Section, we provide a template of a proposal and provide a few tips about how to write a good proposal.

\subsubsection*{Abstract and keywords}

Briefly describe your challenge. Follow the following template (2 sentences maximum each topic):
\begin{itemize}
    \item Background and motivation (stress impact).
    \item Tasks proposed and data used.
    \item Novelty (compared to previous challenges and benchmarks).
    \item Baseline methods and results (positioning the state of the art).
    \item Scientific outcomes expected (list questions asked).
\end{itemize}

Indicate whether this  is a ``regular challenge'' running over a few months, a ``hackathon'' taking place over a day or two, and whether this will include a ``live competition'' in the form of a demonstration requiring on-site presence. Also, provide up to five keywords, from generic to specific.

\subsection*{Competition description}

\subsubsection*{Background and impact}

Provide some background on the problem approached by the competition and fields of research involved. Describe the scope and indicate the anticipated impact of the competition prepared (economical, humanitarian, societal, etc.). Some venues privilege tasks of humanitarian and/or positive societal impact will be particularly considered this year.

Justify the relevance of the problem to the targeted community and indicate whether it is of interest to a large audience or limited to a small number of domain experts (estimate the number of participants). A good consequence for a competition is to learn something new by answering a scientific question or make a significant technical advance.

Describe typical real life scenarios and/or delivery vehicles for the competition. This is particularly important for live competitions, but may also be relevant to regular challenges. For instance: what is the application setting, will you use a virtual or a game environment, what situation(s)/context(s) will participants/players/agents be facing?

Put special emphasis on relating the, necessarily simplified, task of the competition to a real problem faced in industry or academia. If the task cannot be cast in those terms, provide a detailed hypothetical scenario and focus on relevance to the target audience.

Consider adding in a ``hook" as an opening description, to attract those who are unfamiliar with the subject.

\subsubsection*{Novelty}

Have you heard about similar competitions in the past? If yes, describe the key differences.
Indicate whether this is a completely new competition, a competition part of a series, eventually re-using old data.

\subsubsection*{Data}

If the competition uses an evaluation based on the analysis of data, provide detailed information about the available data and
their annotations, as well as permissions or licenses to use such data. If new data are collected or generated, provide details on the procedure, including permissions to collect such data obtained by an ethics committee, if human subjects are involved. In this case, it must be clear in the document that the data will be ready prior to the official launch of the
competition.

Justify that: (1) you have access to large
enough datasets to make the competition interesting and draw
conclusive results; (2) the data will be made freely available after the contest; (3) the ground truth has been kept confidential.

Verify that your dataset is not ``deprecated''. The authors of the original data may have recalled the dataset for some good reasons, e.g. data are biased in some way. The conference you are targeting may supply a list of deprecated datasets. Otherwise searche on the Internet with your dataset name and ``deprecated''. For instance the search for ``tiny images deprecated'' yields this results: ``The deprecation notice for Tiny Images was posted in direct response to a critique by external researchers, who showed that the dataset contained racist and misogynist slurs and other offensive terms, including labels such as {\it rape suspect} and {\it child molester})''. 

Document your dataset thoroughly, using guidelines such as those provided in \cite{Gebru2018dataseheets}. %See Chapter \ref{chap:dataset} for more details on how to prepare a good dataset.

\subsubsection*{Tasks and application scenarios}

Describe the tasks of the competition and explain to which specific real-world scenario(s) they correspond to. If the competition does not lend itself
to real-world scenarios, provide a justification. Justify that the problem posed are scientifically or technically challenging but not impossible to
solve. If data are used, think of illustrating the same scientific problem using several datasets from various application domains.

\subsubsection*{Metrics and evaluation methods}

For quantitative evaluations, select a scoring metric and justify
that it effectively assesses the efficacy of solving the problem
at hand. It should be possible to evaluate the results
objectively. If no metrics are used, explain how the evaluation
will be carried out. Explain how error bars will be computed and/or how the significance in performance difference between participants will be evaluated.

You can include subjective measures provided by human judges (particularly for live /  demonstration competitions). In that case, describe the judging criteria, which must be as orthogonal as possible, sensible, and specific. Provide details on the judging protocol, especially how to break ties between judges. Explain how judges will be recruited and, if possible, give a tentative list of judges, justifying their qualifications.

%See Chapter \ref{chap:judging} for help on evaluating a challenge.

\subsubsection*{Baselines, code, and material provided}

Describe baseline methods that can solve the problems posed in your competition.
Beta-test your competition with such baseline methods and report the results. This is important to demonstrate that the competition is not too easy nor too hard. You should have a range of baseline methods, from simple to sophisticated (state of the art methods). The results should show a large difference between unsophisticated and sophisticated methods.

Make the baseline methods part of the participants' ``starting kit'', which you should make publicly available together with sample data. The starting kit should allow participants to develop their solution and test it in conditions identical to those in which it will be tested on the challenge platform.

For certain competitions, material provided may include a hardware platform. Ideally the participants who cannot afford buying special hardware or do not have access to large computing resources should not be discriminated against. Find a way to make enough resources freely available to deserving participants in need (\eg participant having demonstrated sufficient motivation by going through a screening test).

\subsubsection*{Tutorial and documentation}

Provide a reference to a white paper you wrote describing the
problem and/or explain what tutorial material you will provide. This may include FAQs, Jupyter notebooks, videos, webinars, bootcamps.

\subsection*{Organizational aspects}

\subsubsection*{Protocol}

Explain the procedure of the competition: 
\begin{itemize}
    \item what the participants will have to do, what will be submitted (results or code), and the evaluation procedure;
    \item whether there will there be several phases;
    \item whether you will you use a competition platform with online submissions and a leaderboard;
    \item what you will do for cheating detection and prevention;
    \item what you will do for beta-testing.
\end{itemize}

Code submission competitions can be resource-intensive but offer a plethora of benefits including:
\begin{itemize}
\item A controlled environment.
\item Confidentiality of data.
\item Equal time allocation for participants.
\item Implementation of intricate protocols.
\item Reduced chances of cheating.
\item Accumulation of code for subsequent analysis.
\end{itemize}

\subsubsection*{Rules}

In this section, provide:
\begin{enumerate}
    \item A verbatim copy of (a draft of) the contest rules given to the contestants.
    \item A discussion of those rules and how they lead to the desired outcome of your competition.
    \item A discussion about cheating prevention.
Choose inclusive rules, which allow the broadest possible participation from the target audience.
\end{enumerate}

It is imperative to clearly delineate the rules right from the outset and ensure they remain unchanged throughout. Maintaining transparency is key in fostering trust and participation engagement. Serious competitors prefer well-defined winning conditions. Evaluation procedures must be robust and tested prior to competition launch to prevent issues. Although maintaining consistent rules is vital, organizers should retain the prerogative to amend rules or data if it's deemed essential.
Such modifications, however infrequent, may be necessary to avert nullifying the entire competition. Any alterations made early in the competition are typically more acceptable to participants.
Late-stage changes can cause discontent as participants might've dedicated significant time and resources, and such amendments might nullify their efforts.
Organizers must balance the advantages of a change against its repercussions on the participants. For instance, last-minute minor data corrections might not merit the potential turmoil they could incite amongst competitors.

Organizers face numerous choices like:
\begin{itemize}
  \item The option between single or multiple accounts.
\item Anonymity regulations.
\item Setting limits on submission counts.
\item Deciding between result or code submissions.
\item Instituting rebuttal or review mechanisms for results by fellow participants.  
\end{itemize}

It's beneficial to have an adjudicating body or an uppermost appellate authority.
The winners' codes should be subjected to internal result releases and peer review to ensure authenticity and merit.

We provide a \href{https://codalab.lisn.upsaclay.fr/competitions/3627#learn_the_details-terms_and_conditions}{concrete example of rules}, corresponding to the challenge whose proposal is found in Appendix A.

\subsubsection*{Schedule and readiness}

Provide a timeline for competition preparation and for running the competition itself. Propose a reasonable schedule leaving enough time for the organizers
to prepare the event (a few months), enough time for the participants to develop their methods (e.g. 90 days), enough time for the organizers to review the entries, analyze and publish the results.

For live/demonstration competitions, indicate how much overall time you will need (we do not guarantee all competitions will get the time they request). Also provide a detailed schedule for the on-site contest. This schedule should at least include times for introduction talks/video presentations, demos by the contestants, and an award ceremony.

 Will the participants need to prepare their contribution in advance (e.g. prepare a demonstration) and bring ready-made software and hardware to the competition site? Or, on the contrary, can will they be provided with everything they need to enter the competition on the day of the competition? Do they need to register in advance? What can they expect to be available to them on the premises of the live competition (tables, outlets, hardware, software and network connectivity)? What do they need to bring (multiple connectors, extension cords, etc.)?

Indicate what, at the time of writing this proposal, is already ready.

\subsubsection*{Competition promotion}

Describe the plan that organizers have to promote participation in the competition (e.g., mailing lists in which the call will be distributed, invited talks, etc.).

Also describe your plan  for attracting participants of groups under-represented in competition programs.

\subsection*{Resources}

\subsubsection*{Organizing team}

Provide a short biography of all team members, stressing their competence for their assignments in the competition organization. Please note that diversity in the organizing team is encouraged, please elaborate on this aspect as well.  Make sure to include: coordinators, data providers, platform administrators, baseline method providers, beta testers, and evaluators.

\subsubsection*{Resources provided by organizers, including prizes}

Describe your resources (computers, support staff, equipment, sponsors, and available prizes and travel awards).

For live/demonstration competitions, explain how much will be provided by the organizers (demo framework, software, hardware) and what the participants will need to contribute (laptop, phone, other hardware or software).

\subsubsection*{Support requested}
Indicate the kind of support you need from the conference.

For live/demonstration competitions, indicate what you will need in order to run the live competition. 

\section{A sample successful proposal}

To exemplify the previous guidelines, we provide an example of successful NeurIPS proposal in Appendix A.

\section{Conclusion}

In this chapter, we've covered the fundamentals of organizing competitions. For a more comprehensive understanding of all aspects of competition organization, please refer to the subsequent chapters.

The success of any challenge hinges predominantly on a strong team and a well-structured plan. A few key suggestions: don't underestimate the effort needed to execute your organization plan and bring in additional volunteers as necessary. Allow ample time to beta-test your challenge. If you're new to this, joining a seasoned organizing team to gain hands-on experience is likely the best approach.

\newpage
\section*{Appendix A: Example of competition proposal}
\label{appendixA}

This appendix provides an example of competition proposal, the \href{https://metalearning.chalearn.org/}{Cross-Domain Meta-Learning Challenge}.

\includepdf[pages=-]{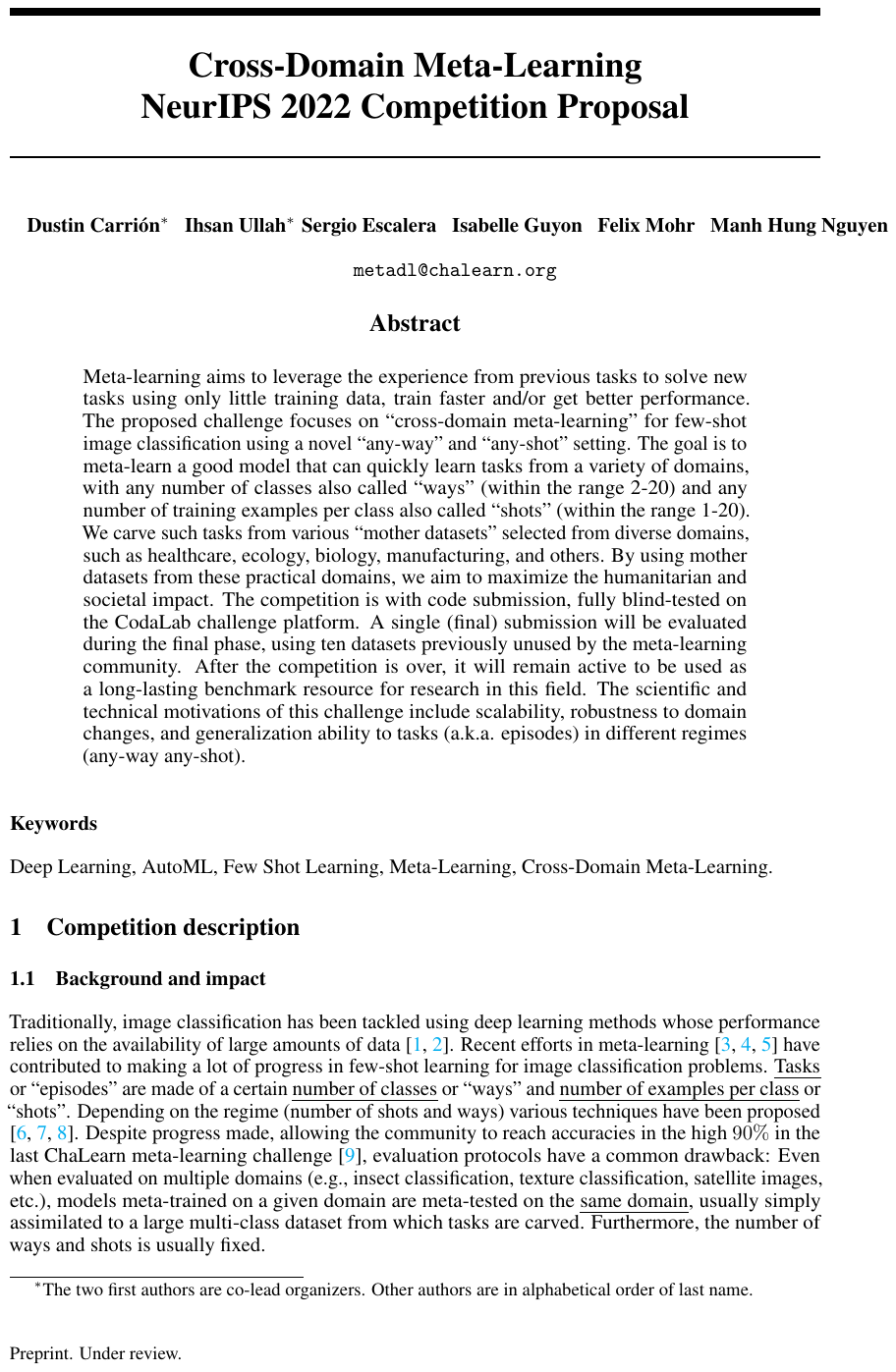}

\section*{Appendix B: Example of fact sheet}
\label{appendixB}

This appendix provides a template of fact sheet, used in the \href{https://metalearning.chalearn.org/}{Cross-Domain Meta-Learning Challenge}. The filled out fact sheets are found on the website of the challenge \footnote{\url{https://metalearning.chalearn.org/}}.

\includepdf[pages=-]{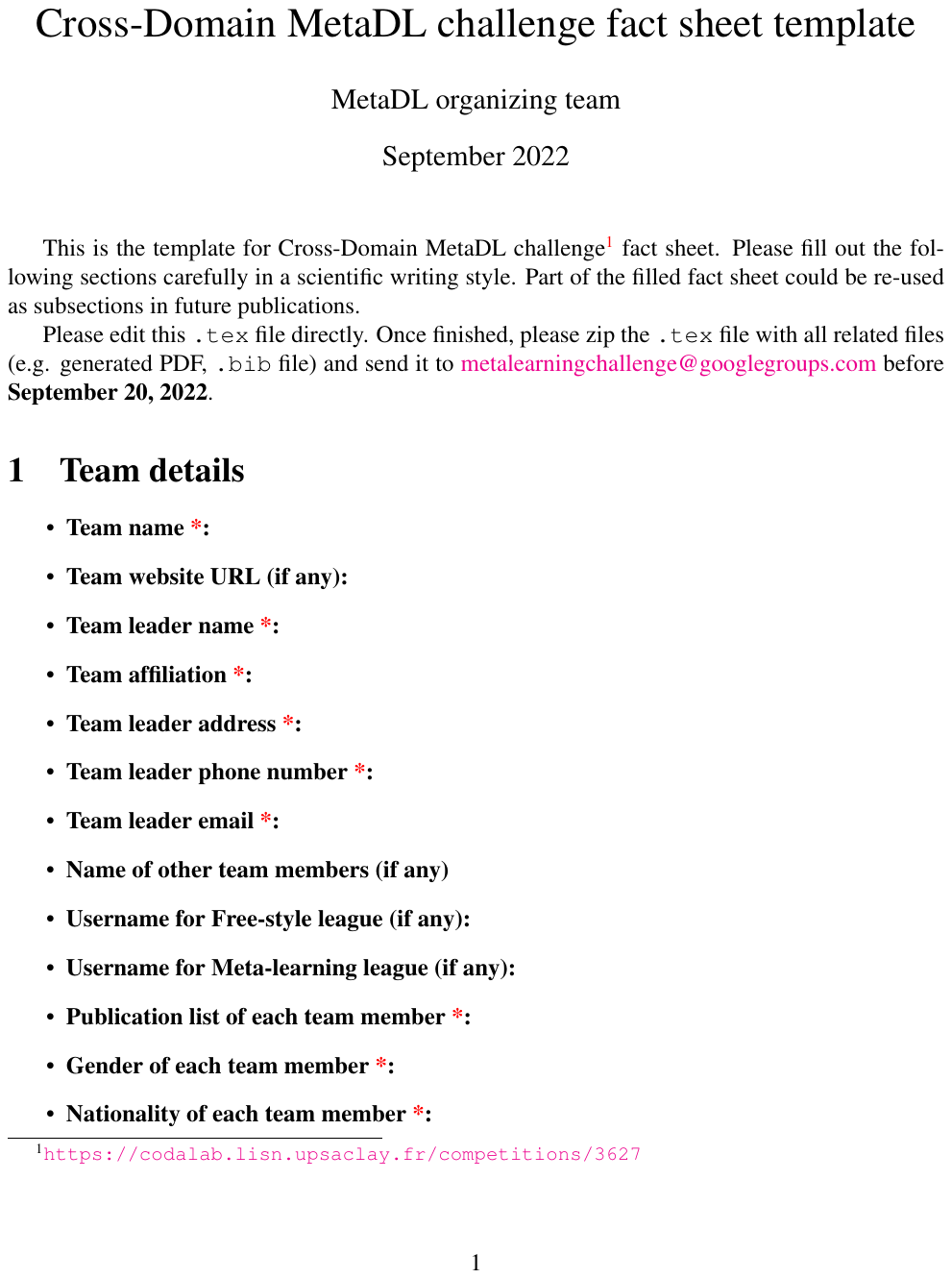}

\section*{Overall challenge design chapter bibliography}

\section*{Acknowledgements: Supported by ANR Chair of Artificial Intelligence HUMANIA ANR-19-CHIA-0022 and TAILOR EU Horizon 2020 grant 952215.}

\bibliographystyle{plain}
\bibliography{ref}

\end{document}